\documentclass[10pt, conference ,compsocconf]{IEEEtran}
\IEEEoverridecommandlockouts
\usepackage{comment}
\usepackage{framed}
    
\usepackage{cite}
\usepackage{amsmath,amssymb,amsfonts}
\usepackage{algorithmic}
\usepackage{graphicx}
\usepackage{textcomp}
\usepackage{xcolor}
\usepackage{url}
\usepackage{multirow}
\usepackage{makecell}
\DeclareMathOperator*{\argmin}{argmin}
\def\BibTeX{{\rm B\kern-.05em{\sc i\kern-.025em b}\kern-.08em
    T\kern-.1667em\lower.7ex\hbox{E}\kern-.125emX}}

\begin{document}

\title{An Analysis of Adversarial Attacks and Defenses on Autonomous Driving Models}

\author{Yao Deng$^{1}$, Xi Zheng$^{1}$, Tianyi Zhang$^{2}$, Chen Chen$^{3}$, Guannan Lou$^{3}$, Miryung Kim$^{4}$   \\ 
  {$^{1}$Macquarie University, 
  Sydney, NSW, Australia} \\
  {$^{2}$Harvard University,
Cambridge, MA, USA} \\
  {$^{3}${University of Sydney, 
Sydney, NSW, Australia}} \\
  {$^{4}${University of California, Los Angeles, CA, USA}} \\
yao.deng@hdr.mq.edu.au, james.zheng@mq.edu.au, tianyi@seas.harvard.edu, \\ \{cche4088, glou2030\}@uni.sydney.edu.au, miryung@cs.ucla.edu
}



\maketitle

\begin{abstract}
Nowadays, autonomous driving has attracted much attention from both industry and academia. 
Convolutional neural network (CNN) is a key component in autonomous driving, which is also increasingly adopted in pervasive computing such as smartphones, wearable devices, and IoT networks.
Prior work shows CNN-based classification models are vulnerable to adversarial attacks. However, it is uncertain to what extent regression models such as driving models are vulnerable to adversarial attacks, the effectiveness of existing defense techniques, and the defense implications for system and middleware builders. 

This paper presents an in-depth analysis of five adversarial attacks and four defense methods on three driving models. Experiments show that, similar to classification models, these models are still highly vulnerable to adversarial attacks. This poses a big security threat to autonomous driving and thus should be taken into account in practice. While these defense methods can effectively defend against different attacks, none of them are able to provide adequate protection against all five attacks. We derive several implications for system and middleware builders: (1) when adding a defense component against adversarial attacks, it is important to deploy multiple defense methods in tandem to achieve a good coverage of various attacks, (2) a black-box attack is much less effective compared with a white-box attack, implying that it is important to keep model details (e.g., model architecture, hyperparameters) confidential via model obfuscation, and (3) driving models with a complex architecture are preferred if computing resources permit as they are more resilient to adversarial attacks than simple models. 
\end{abstract}


\section{Introduction}
Many pervasive computing applications now use regression neural network models. 
For instance, a CNN-based regression model is capable of predicting the distance to collision for unmanned aerial vehicles to accomplish collision-free navigation. A stacked autoencoder regression model is deployed at the edge of simulated sensor networks to predict values of QoS metrics (response time and throughput) for each service~\cite{white2019autoencoders}.
In this paper, we focus
on autonomous driving, which extensively uses CNN-based regression models.

Nowadays, technology companies such as Tesla, Uber, and Waymo have made a huge investment in autonomous vehicles. Waymo recently launched the first self-driving car service in Phoenix, making one of the first steps towards commercializing autonomous vehicles~\cite{waymoS}. 
In an autonomous driving system, cameras and LiDARs are deployed to collect information about the driving scene, which is then fed into a CNN-based driving model to make decisions such as adjusting the speed and steering angle. 

Unfortunately, CNNs can be easily fooled by \textit{adversarial examples}, which are constructed by applying small, pixel-level perturbations to input images~\cite{szegedy2013intriguing,goodfellow2014explaining}. Despite imperceptible to human eyes, such adversarial examples cause CNNs
to make completely wrong decisions. Recently, Tencent Keen Security Lab demonstrated an adversarial attack on Tesla Autopilot by generating adversarial examples to turn on rain wipers when there is no rain~\cite{Tencent}. 

Many adversarial attacks have been proposed and demonstrated effective on image classification models~\cite{szegedy2013intriguing, moosavi2016deepfool,carlini2017towards,xiao2018generating,poursaeed2018generative}. To defend adversarial attacks, several techniques have been proposed to harden neural networks~\cite{goodfellow2014explaining, papernot2016distillation, polytope, huang2017safety}. 
However, previous research mainly focuses on image classification models. It is unclear to what extent these adversarial attacks and defenses are effective on regression models (e.g., autonomous driving models). This uncertainty exposes potential security risks and raises research opportunities. If adversarial attacks could be successfully applied to autonomous driving systems, attackers could easily cause traffic accidents and jeopardize personal 
safety. If existing defense methods cannot be adapted to defend against attacks on regression models, it is imperative to identity a novel defense mechanism suitable for autonomous driving.

This paper presents a comprehensive analysis of five adversarial attack methods and four defense methods on autonomous driving models. By conducting systematic experiments on three driving models, we find that except IT-FGSM ~\cite{kurakin2016adversarialphysical} (36\% attack success rate), all other attacks, including Opt~\cite{szegedy2013intriguing}, Opt\_uni~\cite{moosavi2017universal}, AdvGAN~\cite{poursaeed2018generative}, and AdvGAN\_uni~\cite{poursaeed2018generative}), could effectively generate adversarial examples with an average of 98\% success rate in the white-box setting. Therefore, similar to classification models, CNN-based regression models are also highly vulnerable to adversarial attacks. On the other hand, the attack success rate of all attack methods is significantly lower in the black-box setting (4\% only on average). This implies that, if neural network architecture and hyperparameters are not known, a driving model is much less vulnerable to adversarial attacks. Therefore, in practice, systems and middleware builders should keep their neural networks confidential. It may also be beneficial to apply model obfuscation or model privacy protection techniques~\cite{tramer2016stealing, juuti2019prada}.

In terms of defense, none of the four defense methods can effectively detect all five kinds of attacks. Adversarial training~\cite{goodfellow2014explaining} and defensive distillation~\cite{papernot2016distillation} are only effective to reduce the success rate of two attacks:  IT-FGSM and an optimization based approach. A method that detects abnormal hardware state (e.g., GPU memory usage, GPU utilization rate) can effectively detect these two attacks and to some extent detect AdvGan. Feature squeezing~\cite{xu2017feature}, on the other hand, can detect all five attacks with more than 78\% recall under a specific setting but with a high false positive rate up to 40\%. This indicates that, when building a defense component in a system or middleware, it is necessary to deploy multiple defense methods in tandem to be robust to various attacks. 
Overall, this paper makes the following contributions:
\begin{itemize}
\item{\textbf{Implementation.} 
We replicate five adversarial attack methods and four defense techniques with proper adaptations to cater to regression-based driving models. We release our implementations, models, and datasets for future research and validation.\footnote{Our dataset and models are available at \url{https://github.com/ITSEG-MQ/Adv-attack-and-defense-on-driving-model}}}

\item{\textbf{Evaluation.} 
We comprehensively experiment with five adversarial attacks and four defenses and summarize results from experiments.}

\item{\textbf{Implications.} 
We propose three system building implications for future research in adversarial attacks and defenses on autonomous driving models.}
\end{itemize}

\section{Background and Related Work}
\subsection{Autonomous Driving Model}
\begin{figure}[ht]
\centering
\includegraphics[width = .45\textwidth]{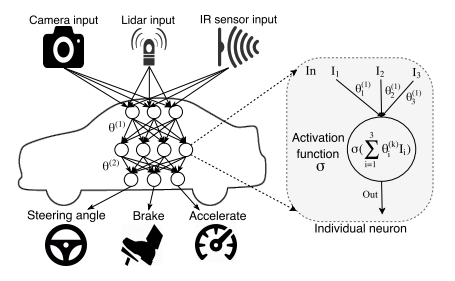}
\caption{The overview of an autonomous driving model \cite{tian2018deeptest}}
\label{driving-model-arch}
\end{figure}

Figure~\ref{driving-model-arch} shows the overview of an autonomous driving model. Given input data from sensors (e.g., LiDARs and cameras), a deep neural network predicts the control of the vehicle such as the steering angle and speed. 
CNN is the mainstream neural network architecture for autonomous driving, since it has excellent performance, requiring less neurons and consuming less resources.
In autonomous vehicles, such driving model is usually included inside a perception domain controller, which can be updated remotely through the vehicle's gateway \cite{pendleton2017perception, liu2017creating}. 
Some companies have published their research on autonomous driving. For example, \textit{comma.ai} presents a CNN based driving model to predict the steering angle based on driving video \cite{santana2016learning}. \textit{Nvidia} builds a CNN model called DAVE-2~\cite{bojarski2016end}. They demonstrate that DAVE-2 can automatically drive a vehicle without human intervention 90\% of the time in a simulation test while performing autonomous steering 98\% of time in an on-road test. 
\subsection{Adversarial Attacks}
By adding small perturbations to original images, adversarial attacks can deceive a target model to produce completely wrong predictions. 
Currently, adversarial attacks are mainly researched on image classification tasks. Given a target model \(f\) and an original image \(x \in \mathcal{X}\) with its class \(c\), an adversarial attack constructs an imperceptible adversarial  perturbation \(\delta\) to form an adversarial example \(x' = x + \epsilon \) and make the target model classify \(x'\) as \(c'\) that is different from \(c\). 

Depending on the information required to perform the attack, existing adversarial attacks can be categorized into {\em white-box attacks} and {\em black-box attacks}. White-box attacks require all details of a target model to perform the attack, including the training data, the neural network architecture, parameters, and hyper-parameters, as well as the privilege to gather the gradients and prediction results of a model \cite{yuan2019adversarial}. By contrast, black-box attacks only require to query the model with arbitrary input data and get the prediction result \cite{papernot2017practical}. Based on the inputs and outputs from the target model, to perform a black-box attack, attackers can build a substitute model and achieve white-box attacks on their own model. The adversarial examples on the substitute model could then be used to attack the target black-box model, which is called the {\em transferability} of adversarial examples. 

\subsection{Adversarial Defenses}
Several techniques have been proposed to defend adversarial attacks, which can be roughly categorized as proactive defenses and reactive defenses.
Proactive defenses aim to improve the robustness of a neural network against adversarial examples. The common method is to retrain the model using a dataset with adversarial examples~\cite{goodfellow2014explaining}, or to add regularization components to the target model~\cite{yan2018deep}. Furthermore, Papernot et al.~introduced {\em defensive distillation}. It increases the magnitude of inputs to the Softmax layer by adjusting a parameter $T$ called {\em Temperature} 
to harden a neural network~\cite{papernot2016distillation}.
Reactive methods, on the other hand, aim to detect adversarial examples. DNN could be used to determine whether the model is under attack, by verifying the properties of input images~\cite{zheng2018robust} or by detecting the status of the Softmax layer~\cite{lee2018simple}.

\section{Methodology}
\subsection{Adversarial Attacks on Autonomous Driving Models}
\label{sec:attack-details}

 For an image classification model, an adversarial attack is considered successful if an adversarial image is classified as a different class compared with the original image. However, autonomous driving models are regression models that predict 
 continuous values.
Therefore, adversarial attacks on driving models are defined with respect to an acceptable error range, known as {\em adversarial threshold}. 
Hence, an adversarial attack on a driving model is considered successful if the deviation between the original prediction and the prediction of an adversarial example is above the adversarial threshold. 

Current adversarial attacks on classification model could be categorized into three classes based on the perturbation generation method. Fast Gradient Sign based method~\cite{goodfellow2014explaining} directly generates adversarial examples by adding the sign of the loss gradient with respect to each pixel on original images. Optimization-based method~\cite{szegedy2013intriguing} formulates the adversarial example construction as an optimization problem. Generative model based method~\cite{poursaeed2018generative} 
proposes to generate adversarial examples by harnessing the power of generative models such as Generative Adversarial Network (GAN)~\cite{goodfellow2014generative} and autoencoder  networks~\cite{baldi2012autoencoders}.  In addition, there is a special attack named {\em universal attack}~\cite{moosavi2017universal}, which generates a single adversarial example to fail all samples in the dataset.

In this study, we re-implement five adversarial attacks to form a comprehensive set of adversarial attacks on regression models. 
We first choose two classic adversarial attacks: {\em Iterative Targeted Fast Gradient Sign Method} (IT-FGSM)~\cite{kurakin2016adversarialphysical, kurakin2016adversarial}, a variant of the classic method Fast Gradient Sign Method (FGSM)~\cite{goodfellow2014explaining}, and an optimization-based approach~\cite{szegedy2013intriguing} as it is the first approach to generate adversarial examples. We then choose a state-of-the-art generative model based attack called AdvGAN~\cite{poursaeed2018generative}. Furthermore, we implement two universal attack methods to increase the diversity of attacks in our experiments. We do not choose attack methods such as C\&W attack~\cite{carlini2017towards} and DeepFool~\cite{moosavi2016deepfool}, since these attacks rely on the attributes of classification models (e.g. decision boundary and the Softmax function). Thus, they cannot be adapted to regression models. We elaborate on the five selected attack methods below. 

\subsubsection{Iterative Targeted Fast Gradient Sign Method (IT-FGSM)} IT-FGSM~\cite{kurakin2016adversarialphysical, kurakin2016adversarial} is a variant of Fast Gradient Sign Method (FGSM) that simply adds the sign of the loss gradient with respect to each pixel on original images. IT-FGSM applies the targeted FGSM multiple times to get a more powerful adversarial example.

\subsubsection{Optimization Based Approach (Opt)} This approach calculates an adversarial perturbation \(\epsilon\) for classification models by solving the optimization problem as in Formula (\ref{lbfgs})~\cite{szegedy2013intriguing}.
\begin{equation}
\label{lbfgs}
\argmin_{\epsilon}||\epsilon||_2  \hspace{3mm}s.t. \hspace{2mm}f(x+\epsilon)=c',x+\epsilon \in[0,1]^m
\end{equation}

For regression models, we change \(c'\) to \(f(x) + \Delta\) and adapt Formula (\ref{lbfgs}) to Formula (\ref{opt-based}) and apply the Adam optimizer~\cite{kingma2014adam} to solve the optimization problem.
\begin{equation}
\label{opt-based}
\argmin_{\epsilon}\hspace{1mm}||\epsilon||_2+J_{\theta}(Clip(x+\epsilon),f(x) + \Delta)
\end{equation}

\subsubsection{AdvGAN}
AdvGAN generates an adversarial example \(\mathcal{G}(x)\) from an original image by integrating another objective \(\mathcal{L}_y = J_\theta(\mathcal{G}(x), f(x) + \Delta)\) into the objective function \(\mathcal{L}_{AdvGAN} = \mathcal{L}_y + \alpha \mathcal{L}_{GAN}\), where \(\alpha\) sets the importance of each objective. After training, the generator \(\mathcal{G}\) could generate an adversarial example \(x'\) that is similar to an original image but make an prediction \(f(x')\) that deviates \(\Delta\) from \(f(x)\).
    

\subsubsection{Universal Adversarial Perturbation (Opt\_uni)}
We implement this attack based on the optimization based approach. We first generate a perturbation \(v\) on one image. Then for each image in the dataset, we calculate the minimal change \(\Delta v\) and adapt \(v\) to \(v + \delta v\). When iterating the whole dataset, an universal perturbation is obtained. 

\subsubsection{AdvGAN Universal Adversarial Perturbation (AdvGAN\_uni)}
Poursaeed et al.~proposed to use GAN to generate universal adversarial perturbations~\cite{poursaeed2018generative}. Instead of using the generator to construct an unique perturbation \((\mathcal{G}(x))\) for each input image, the generator outputs a universal adversarial perturbation. In this study, we implement this approach based on the AdvGAN architecture.

\subsection{Adversarial Defenses on Autonomous Driving Models}
\label{sec:adv_defense}

While there is a proliferation of defense methods against adversarial attacks, many of them are designed for image classification tasks only and thus are not applicable to regression based autonomous driving models. For example, the adversarial transformation method \cite{guo2017countering} reduces an adversarial attack success rate by randomly clipping and rotating input images. Rotating input images
may not impact an image classification model. But for a regression driving model, rotating input images will cause a big prediction error. Adversarial de-noiser~\cite{shen2017ape} also has the same problem, since it has to apply transformations on input images. Therefore, there is a need for reevaluating and designing defense methods for regression models and adapting them to autonomous driving. 

In this paper, we adapt and re-implement four defense methods for autonomous driving. First, we choose two classic proactive defense methods, \textit{adversarial training}~\cite{goodfellow2014explaining} and \textit{defensive distillation}~\cite{papernot2016distillation}. Then we develop a reactive defense method based on the insight that real-time adversarial example generation may lead to a resource usage spike in autonomous vehicles. This method performs runtime monitoring in autonomous vehicles and detects anomalous hardware usage rates. 
Finally, we choose another state-of-the-art adversarial attack detection method called \textit{feature squeezing}~\cite{xu2017feature}, since it does not need an auxiliary model and achieves good performance on classification models.

\subsubsection{Adversarial Training}
By retraining the original model with adversarial examples, the new model learns features of adversarial examples and thus has better generalization and robustness. In this study, we add adversarial examples generated by proposed attacks to train a new model. 

\subsubsection{Defensive Distillation}
Defensive distillation~\cite{papernot2016distillation} uses class probabilities predicted by the original model as soft labels to train a new model. We adapt the original defensive distillation approach to handle regression models based on the finding by Hinton et al.~\cite{hinton2006reducing}. They show that the output of a neural network hidden layer contains highly encoded information that can be leveraged for model distillation. Similarly, the output of fully connected layers can be used to perform defensive distillation on CNN based driving models. We observe that for inputs \(x_i\) that have similar outputs \(f(x_i)\), there are indeed uncorrelated features in the last fully connected layer outputs \(z_i^d\), which can provide additional information to distill and train a new model \(g\). While training the distilled model, we add a regularizer to distill the information from the original model as shown in Formula~(\ref{distillation_r}). In this way, we use information from both the output \(f(x)\) and the tensors \(z^d\) to train the distilled model in order to enhance its generalization and robustness against adversarial attacks. 

\begin{equation}
\label{distillation_r}
\mathcal{L} = \Sigma_1^n(\lambda(||z_i^d - {z'_i}^d||) + ||f(x_i) - g(x_i)||)/n
\end{equation}

\subsubsection{Anomaly Detection}

Autonomous vehicles are usually equipped with a runtime monitoring system to check the vehicle state~\cite{watanabe2018runtime, zheng2018efficient}. First, 
we monitor model prediction latency caused by adversarial attacks. Second, since autonomous vehicles are resource constrained, we monitor any spikes in GPU memory usage and GPU utilization rate via Nvidia System Management Interface (Nvidia-smi) to detect additional computation caused by adversarial attacks. 
We evaluate the effectiveness of this anomaly detection approach by comparing the prediction time per image and GPU usage with and without adversarial attacks and further investigate its effectiveness on different kinds of attack methods. 

\subsubsection{Feature squeezing}
Xu et al.~\cite{xu2017feature} proposed two feature squeezing methods for adversarial defense. The first method squeezes the original 24-bit color down to 1 bit to 8 bit color. By doing so, adversarial noise becomes more perceptible as the bit depth decreases. The second method adopts median spatial smoothing, which moves a filter move across an original image and modifies the center pixel value to the median of the pixel values in the filter. 
If the difference between the prediction result of the original image and the prediction result of a squeezed image by either of the two methods exceeds a pre-defined threshold \(T\), then the given input is likely to be an adversarial example.

\section{Experiment}
\noindent{\bf\em Dataset.} We use the Udacity dataset \cite{Udacity} to train three autonomous driving models and generate adversarial examples. This dataset contains real-world road images collected by a front camera installed in a vehicle and the dataset is splited to training set and testing set by Udacity.   The training set contains $33805$ frames and the test set contains $5614$ frames. 
The steering angle of each frame is normalized from a degree to a value range between $-1$ and $1$.
\noindent{\bf\em Autonomous driving models.} We implement and train three driving models, Epoch \cite{epoch}, Nvidia DAVE-2 \cite{bojarski2016end}, and VGG16 \cite{simonyan2014very} using Pytorch. We choose Epoch because it performs well in the Udacity Challenge~\cite{Udacity}. Nvidia DAVE-2 is a well known, publicly autonomous driving model. VGG16 adopts a highly robust neural network architecture that is widely used in transfer learning for image classification. We adapt VGG16 to a regression driving model by replacing its last layer with a three-layer feed-forward network.
For these driving models, we uniformly set their input image size as $128*128$. The details of those models are demonstrated in Table~\ref{models}. The error rate of these models is measured by Root Mean Square Error, as shown in column {\sf RMSE}. 
On the test dataset, if an autonomous driving model by default predicts $0$ for all frames, the RMSE between the predictions and the ground truth is $0.20678$. The RMSE of our adapted VGG16, Epoch and Nvidia DAVE-2 are $0.0906$, $0.0962$ and $0.1055$ respectively. These models would be ranked 6th to 8th in the Udacity leaderboard, implying that they are fairly accurate on the Udacity test set.
The driving models with higher rankings are all based on more complicated neural network architectures (e.g., CNN+RNN), which 
might be
less
susceptible to adversarial attacks and we plan to investigate in future work. 

\begin{table}[]
\centering
\caption{Three autonomous driving models for experiment}
\label{models}
\renewcommand{\arraystretch}{1.2}
\renewcommand\tabcolsep{5pt}
\begin{tabular}{|c|r|r|r|}
\hline
\textsf{Model}  & \textsf{Parameters\#} & \textsf{Size(MB)} & \textsf{RMSE} \\ \hline
Epoch  & $33,649,729$            & $147.82$            & $0.0962$        \\ \hline
Nvidia & $6,288,765$             & $26.76$             & $0.1055$        \\ \hline
VGG16  & $68,246,337$            & $332.29$            & $0.0906$        \\ \hline
\end{tabular}
\end{table}

\begin{figure}[t]
\centering
\includegraphics[width = .4\textwidth]{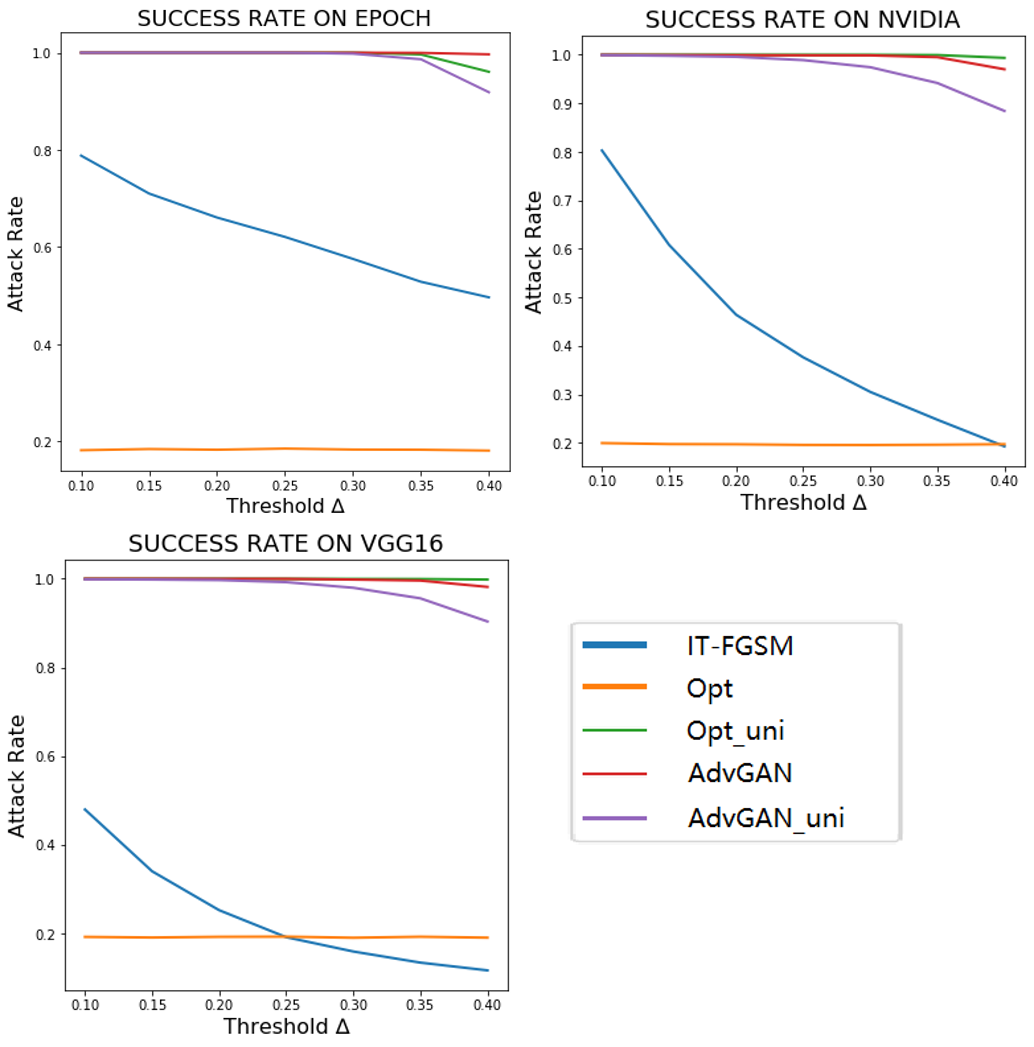}
\caption{Attack Success Rate with Different Threshold \(\Delta\) }
\label{delta_change}
\end{figure}

\begin{figure}[t]
\centering
\includegraphics[width = .4\textwidth]{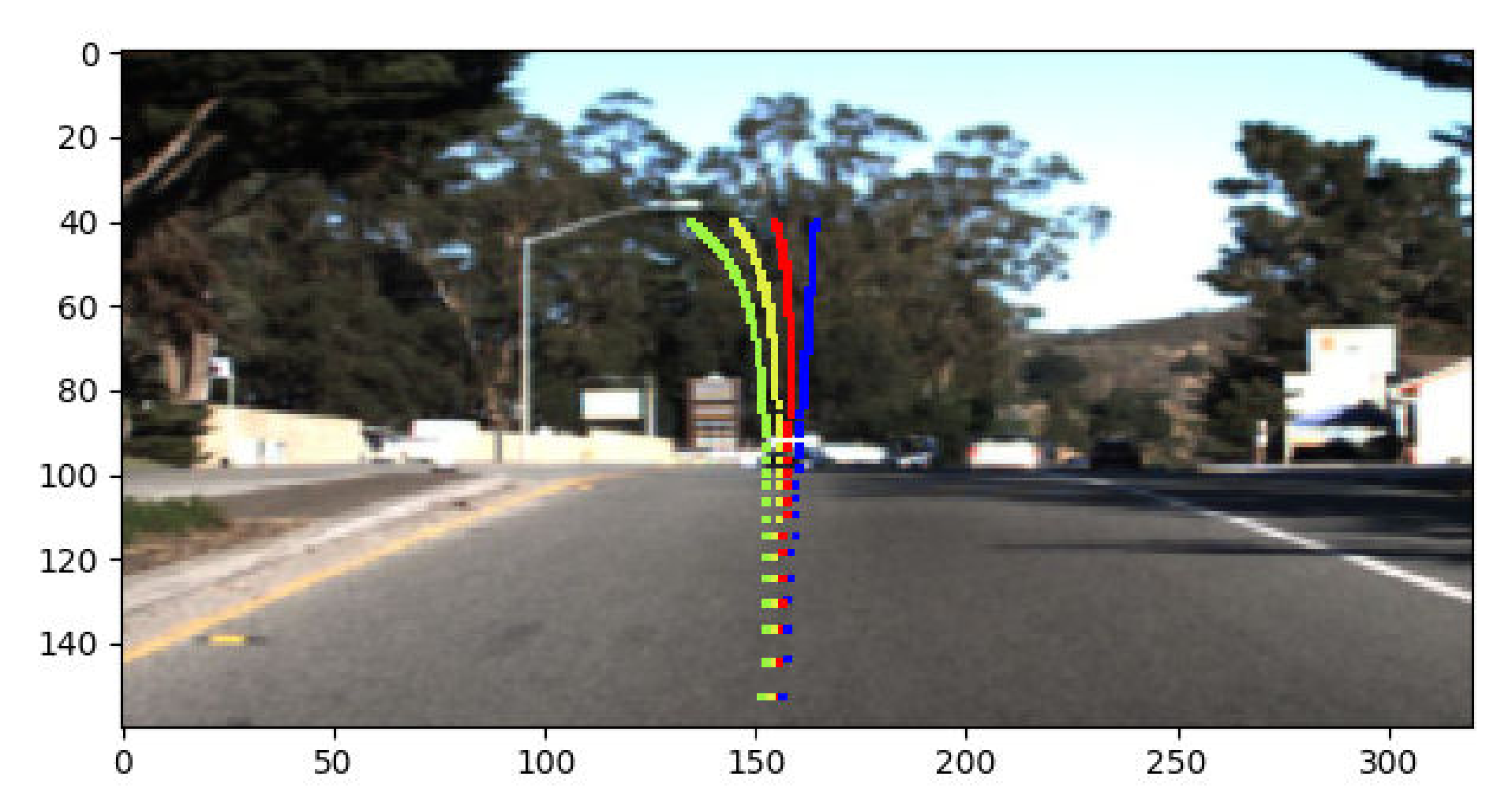}
\caption{Display of steering angle track on the driving scene image. The blue, red, yellow and green lines are the tracks with 0, 0.1, 0.2 and 0.3 steering angle deviations respectively. When setting the deviation to 0.3, the vehicle clearly turns to a wrong direction.}
\label{deviation}
\end{figure}

\noindent{\bf\em Adversarial attack and defense settings}
All attacks and defenses are implemented in Pytorch. We set the adversarial threshold \(\Delta\) to 0.3. In other words, we consider the attack successful if the difference between the steer angle prediction on an adversarial example and the original prediction is greater than 0.3. 
Figure~\ref{delta_change} shows the impact of threshold on attack success rate. With the increase of the threshold, the limitation on the attack success rate becomes strict, and the attack success rate drops down.
When the threshold is less than 0.3, the success rate of the universal perturbation attack and AdvGAN attack keeps steady. When the threshold reaches to 0.3, the success rate of 
all five attack methods starts to decrease. And according to Figure~\ref{deviation}, steering angle deviation achieving 0.3 could cause a significant cumulative displacement. Thus, 0.3 is selected as the threshold in this study, and the threshold could be adjusted according to actual needs in the real-world implement.



For Iterative Targeted Fast Gradient Sign Method (IT-FGSM), we set the perturbation control parameter \(\epsilon\) to 0.01 and the iteration number to 5. For the optimization-based approach, we set the learning rate of Adam optimizer to 0.005 and the maximum iteration number to 100 to force the algorithm to stop when it cannot find the optimal solution in reasonable time. For the universal adversarial perturbation attack, we keep the same setting with the optimization-based approach. For AdvGAN, we set the learning rate of Adam optimizer of the discriminator and the generator to 0.001. We clip the value of perturbation in the range [-0.3, 0.3]. We set \(\alpha\) as 1. For adversarial training, we set \(\alpha\) as 0.5. For defensive distillation, we vary \(\lambda\) as $0.01$, $0.05$, $0.1$, $0.5$, $1$, $5$, $10$ respectively to train $7$ different distillation 
models and experiment with their performance. All hyper-parameter settings either use the default ones or the same values from prior work. For feature squeezing, we implement 4-bit image depth reduction and \(2\times2\) median smoothing because they perform best as shown in~\cite{xu2017feature}. We vary the threshold as $0.01$, $0.05$, $0.1$ and $0.15$ to explore performances of feature squeezing under different settings.

We conduct experiments under white-box and black-box settings. Under white-box setting, all five attacks have full knowledge of a driving model so they could directly generate adversarial examples by malware. Under black-box setting, attakers are assumed to train a proxy driving model offline. For universal perturbation attacks and AdvGAN universal adversarial perturbation attacks, attackers could construct universal perturbations based on a proxy driving model. For AdvGAN attack, the AdvGAN model could also be trained on the proxy model. Then universal perturbations and AdvGAN could be integrated into a malware to conduct black-box attacks. Other two attacks are not available under the black-box setting as they need information about the driving model in real-time to construct adversarial examples. Therefore, we only evaluate Universal perturbation attack, AdvGAN attack and AdvGAN universal perturbation attack in black-box attack experiments. 

We use a common metric, attack success rate, to measure the performance of adversarial attacks and defenses. 
An attack is considered successful, if the steer angle deviation is greater than the adversarial threshold \(\Delta\) = $0.3$. An attack success rate is computed as the portion of the number of adversarial examples that the attack is successful out of all generated adversarial examples. All experiments are conducted in a simulation environment with Intel i7-8700 3.2GHz, 32GB of memory, and a NVIDIA RTX 2080Ti GPU.


\subsection{Research Questions}
We investigate the following research questions:
\begin{itemize}
\item {\bf RQ1:} In the white-box setting, how does each of the five attacks perform on different driving models?
\item {\bf RQ2:} Do those attacks still perform well in the black-box setting? 
\item {\bf RQ3:} Do adversarial defense methods improve the robustness of  driving models against adversarial attacks? 
\end{itemize}

\section{Result}

\begin{figure*}[!t]

\centering
\includegraphics[width = .85\textwidth]{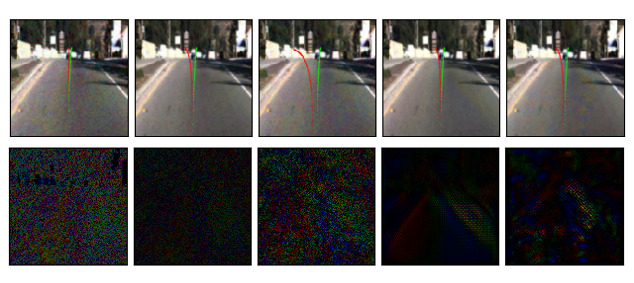}
\caption{The comparison of five adversarial attacks on a road image. The top row shows the adversarial images generated by five attacks (from left to right: IT-FGSM, Opt, Opt\_uni, AdvGAN, AdvGAN\_uni), as well as the steer angle predictions of the Epoch model on the original image (green line, value is -0.0423) and on the adversarial image (red line, values are 0.0308, 0.2665, 0.6737, 0.3649 and 0.4658 respectively). The bottom row shows the perturbation of each attack amplified by five times.}
\label{exp1-3}
\end{figure*}


This section presents experiment results of five attacks and three defenses on three driving models. 

\begin{table}[]
\caption{Attack Success Rate of five attacks on three driving models}
\label{exp1-1}
\centering
\renewcommand{\arraystretch}{1.2}
\renewcommand\tabcolsep{5pt}
\resizebox{0.65\columnwidth}{!}{
\begin{tabular}{|l|r|r|r|}
\hline
                 & Epoch  & Nvidia & VGG16  \\ \hline
IT-FGSM          & 59.2\% & 31.8\% & 16.2\% \\ \hline
Opt     & 91.2\% & 99.3\% & 95.8\% \\ \hline
Opt\_uni        & 100\%    & 99.9\% & 99.9\% \\ \hline
AdvGAN           & 100\%    & 99.5\% & 98.0\% \\ \hline
AdvGAN\_uni & 99.8\% & 96.4\% & 96.3\% \\ \hline
\end{tabular}
}
\end{table}

\subsection{RQ1. Effectiveness of White-box Attacks}
Table \ref{exp1-1} shows the attack success rate of applying five attacks on three selected driving models. All attack methods except IT-FGSM achieve high attack success rate on all three models. The highest attack success rate of IT-FGSM is 59.2\% on Epoch model but other methods can achieve over 90\% on all three models. Opt\_uni and AdvGAN even achieve 100\% on Epoch model and average over 99\% on the other two models. The reason is that IT-FGSM only perturbs pixels in an image by simply adding the sign of gradients, while Opt and Opt\_uni utilize Adam optimizer to search adversarial perturbations in multiple iterations. AdvGAN and AdvGAN\_uni learn intrinsic features (e.g., yellow road lanes) that influence steer angle predictions to generate adversarial perturbations. IT-FGSM has the lowest attack success rate on VGG16, implying that the driving model with a more complicated structure is robust to simple adversarial attacks. However, for other attacks, attack success rates are fairly high on three models. The result indicates that autonomous driving models are vulnerable to adversarial attacks. 

Figure~\ref{exp1-3} illustrates steer angle deviations predicted by Epoch. Under different kinds of attacks (the top row), as well as corresponding adversarial perturbations generated on the same driving scene (the bottom row). IT-FGSM causes only a slight steer angle deviation (0.0731 only) even though this attack adds a perturbation with a big distortion, while other attacks cause much larger deviations close to or above the adversarial threshold (0.3). The amplified perturbations generated by AdvGAN and AdvGAN\_Uni resemble the yellow lane and the white lane in the original image, implying that these GAN-based attacks have learned that road lanes are important features to affect the steer angle prediction.

In summary, adversarial attacks are achievable and dangerous on autonomous driving. Optimization-based attacks (Opt, Opt\_uni) and Generative-network based attacks (AdvGAN, AdvGAN\_uni) could generate adversarial examples to achieve high attack success rate under white-box setting. Among all the attacks, AdvGAN seems to be the most dangerous attack as it exploits intrinsic features learned by driving models.

\begin{framed}
\noindent Result 1:  Regression driving models are vulnerable to adversarial attacks as IT-FGSM, Opt, Opt\_uni, AdvGAN, and AdvGAN\_uni all achieve high attack success rates. 

\end{framed}

\begin{table}[!h]
\centering
\caption{Attack Success Rate under Black-box Attacks}
\label{exp2}
\renewcommand{\arraystretch}{1.2}
\renewcommand\tabcolsep{5pt}
\scalebox{1}{

\begin{tabular}{|l|l|c|c|c|}
\hline
                        &             & \multicolumn{1}{l|}{Epoch} & \multicolumn{1}{l|}{Nvidia} & \multicolumn{1}{l|}{VGG16} \\ \hline
\multirow{3}{*}{Epoch}  & Opt\_uni    & -                          & 5.4\%                       & 0.5\%                      \\ \cline{2-5} 
                        & AdvGAN      & -                          & 0.1\%                       & 0.2\%                      \\ \cline{2-5} 
                        & AdvGAN\_uni & -                          & 3.7\%                       & 0.7\%                      \\ \hline
\multirow{3}{*}{Nvidia} & Opt\_uni    & 9.8\%                      & -                           & 2.6\%                      \\ \cline{2-5} 
                        & AdvGAN      & 2.5\%                      & -                           & 0.4\%                      \\ \cline{2-5} 
                        & AdvGAN\_uni & 6.4\%                      & -                           & 8.4\%                      \\ \hline
\multirow{3}{*}{VGG16}  & Opt\_uni    & 3.9\%                      & 30.0\%                      & -                          \\ \cline{2-5} 
                        & AdvGAN      & 0.6\%                      & 0.1\%                       & -                          \\ \cline{2-5} 
                        & AdvGAN\_uni & 1.3\%                      & 1.7\%                       & -                          \\ \hline
\end{tabular}
}
\end{table}

\begin{table*}[!t]
\caption{Attack Success Rate with Adversarial Training}
\label{exp3-1}
\centering
\renewcommand{\arraystretch}{1.3}
\renewcommand\tabcolsep{3pt}
\begin{tabular}{|c|c|r|r|r|r|r|r|}
\hline
\multirow{2}{*}{{\sf Model}} & \multirow{2}{*}{{\sf Data Augmentation}} & \multirow{2}{*}{{\sf RMSE}} & \multicolumn{5}{c|}{{\sf Attack Success Rate}} \\ \cline{4-8} 
\multicolumn{1}{|c|} {} & {} &  & IT-FGSM & Opt & Opt\_Uni & AdvGAN & AdvGAN\_uni \\ \hline
\multirow{6}{*}{\begin{tabular}[c]{@{}c@{}}Epoch\end{tabular}} & --- & 0.0962 & 59.2\% & 91.2\% & 99.9\% & 99.5\% & 99.8\% \\ \cline{2-8} 
 & IT-FGSM & 0.0889 & 39.3\% & 95.2\% & 100\% & 99.9\% & 99.2\% \\ \cline{2-8} 
 & Opt & 0.1029 & 53.2\% & 90.3\% & 100\% & 99.9\% & 99.1\% \\ \cline{2-8} 
 & Opt\_Uni & 0.0953 & 47.3\% & 94.8\% & 100\% & 100\% & 99.6\% \\ \cline{2-8} 
 & AdvGAN & 0.1014 & 52.2\% & 96.8\% & 100\% & 99.8\% & 99.2\% \\ \cline{2-8} 
 & AdvGAN\_uni & 0.0888 & 48.4\% & 96.3\% & 100\% & 99.9\% & 99.3\% \\ \Xhline{1.2pt}
\multirow{6}{*}{\begin{tabular}[c]{@{}c@{}}Nvidia\end{tabular}} & --- & 0.1055 & 31.8\% & 99.3\% & 99.9\% & 99.5\% & 96.4\% \\ \cline{2-8} 
 & IT-FGSM & 0.1119 & 12.0\% & 98.8\% & 100\% & 99.7\% & 94.7\% \\ \cline{2-8} 
 & Opt & 0.1119 & 16.0\% & 90.3\% & 100\% & 99.7\% & 96.3\% \\ \cline{2-8} 
 & Opt\_Uni & 0.1144 & 11.2\% & 92.1\% & 100\% & 99.7\% & 94.0\% \\ \cline{2-8} 
 & AdvGAN & 0.1118 & 15.0\% & 99.6\% & 100\% & 99.8\% & 93.5\% \\ \cline{2-8} 
 & AdvGAN\_uni & 0.1162 & 13.3\% & 92.3\% & 100\% & 99.8\% & 89.5\% \\ \Xhline{1.2pt}
\multirow{6}{*}{\begin{tabular}[c]{@{}c@{}}VGG16\end{tabular}} & --- & 0.0906 & 16.2\% & 95.8\% & 99.9\% & 98.0\% & 96.3\% \\ \cline{2-8} 
 & IT-FGSM & 0.0889 & 5.8\% & 99.1\% & 99.9\% & 99.0\% & 97.3\% \\ \cline{2-8} 
 & Opt & 0.0893 & 7.3\% & 100\% & 100\% & 99.0\% & 96.8\% \\ \cline{2-8} 
 & Opt\_Uni & 0.0875 & 6.1\% & 98.7\% & 100\% & 98.8\% & 97.2\% \\ \cline{2-8} 
 & AdvGAN & 0.0788 & 3.1\% & 96.8\% & 100\% & 99.8\% & 99.2\% \\ \cline{2-8} 
 & AdvGAN\_uni & 0.0922 & 4.5\% & 99.2\% & 100\% & 98.3\% & 91.8 \% \\ \hline
\end{tabular}
\end{table*}

\subsection{RQ2. Effectiveness of Black-box Attacks}
\label{sec:rq2}
To investigate the effectiveness of different adversarial attacks in the black-box setting, we first generate adversarial examples on each model and then reuse these examples to attack the other models.


Table \ref{exp2} shows the attack success rate when reusing adversarial examples across models. Overall, the attack success rate of Opt\_uni, AdvGAN, AdvGAN\_uni drops significantly in the black-box setting, compared within the white-box setting. 
Opt\_Uni is relatively more effective than other attacks in the black-box setting. For instance, adversarial examples generated by Opt\_Uni on VGG16 achieves 30.0\% success rate on the Nvidia model. The reason may be that perturbations 
generated by Opt\_Uni lead to bigger image distortions than other attacks, which have higher chances to cause prediction deviation in the black-box setting. However, adversarial perturbations generated by other attacks are specialized towards individual models and thus lose the advantages when reused across models.

Furthermore, adversarial examples generated on the most complicated driving model VGG16 have better transferability than those on
the other two simpler models. For instance, adversarial examples generated by Opt\_uni on VGG16 could achieve a 30.0\% attack success rate on Nvidia while adversarial examples generated by the same attack on Epoch could only achieve 5.4\% success rate.
VGG16 is also more robust against black-box attacks than the other two models. For example, adversarial examples generated on the Nvidia model by Opt\_uni achieve a 9.8\% attack success rate on Epoch while only 2.6\% on VGG16. These results indicate that the transferability of adversarial examples may relate to intrinsic properties of driving models such as the complexity of network architecture, which should be verified in further research.

Prior works show that adversarial examples generated on a classification model could be used to successfully attack other models with different architectures for the same task~\cite{liu2016delving, papernot2016transferability, papernot2017practical}. However, our experiment is inconsistent with those previous findings of the transferability of adversarial attacks on classification models. 
It implies that adversarial attacks may have different attributes on regression models. 
In~\cite{goodfellow2014explaining}, authors propose a new explanation for transferability. They find out that for the same classification task, the models with different architecture would learn a similar function and decision boundary so that the adversarial examples generated on one model could attack the others.
For regression models, the conflicting result from the above findings may suggest that different regression models fit in different hyper-planes so that the adversarial examples generated on one model cannot be transferred to attack other models. 

\begin{framed}
\noindent Result 2:  Attacks under black-box setting do not perform well on autonomous driving models. The result demonstrates that the transferability of adversarial examples for driving models is not good, which contradicts previous experiment results on classification models. 
\end{framed}

\subsection{RQ3. Effectiveness of Adversarial Defenses}

\begin{figure}[!t]
\centering
\includegraphics[width = 0.4\textwidth]{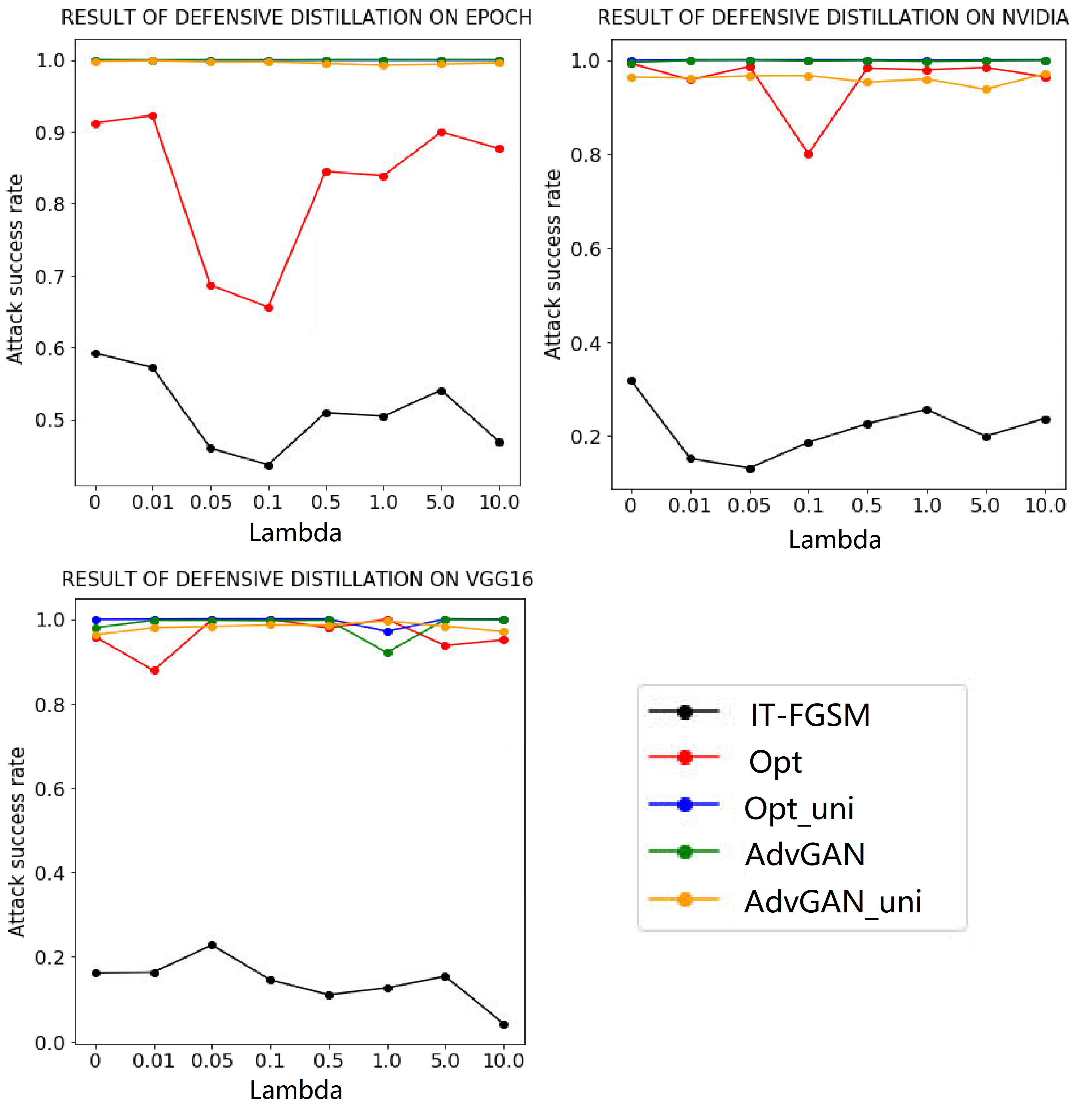}
\caption{Attack Success Rate with Defensive Distillation}
\label{exp3-2}
\end{figure}

\begin{table}[!t]
\centering
\caption{Adversarial attack overhead on driving models}
\label{exp3-3}
\renewcommand{\arraystretch}{1.3}
\renewcommand\tabcolsep{3pt}
\scalebox{0.9}{
\begin{tabular}{|c|c|r|r|r|}
\hline
\multicolumn{2}{|c|}{} & \begin{tabular}[c]{@{}c@{}}{\sf Prediction Time} \\ {\sf Overhead (s)}\end{tabular} & \begin{tabular}[c]{@{}c@{}}{\sf GPU Memory} \\ {\sf Overhead (\%)}\end{tabular} & \begin{tabular}[c]{@{}c@{}}{\sf GPU Utilization}\\{\sf Overhead (\%)}\end{tabular} \\ \hline
\multirow{5}{*}{\begin{tabular}[c]{@{}c@{}}Epoch\end{tabular}} & IT-FGSM & 0.04243 & 50.62 & 36 \\ \cline{2-5} 
 & Opt & 0.10490 & 50.91 & 32 \\ \cline{2-5} 
 & Opt\_uni & -0.00007 & 0.38 & 1 \\ \cline{2-5} 
 & AdvGAN & 0.00532 & 4.01 & 1 \\ \cline{2-5} 
 & AdvGAN\_uni & -0.00013 & 0.19 & 1 \\ \hline
\multirow{5}{*}{\begin{tabular}[c]{@{}c@{}}Nvidia\end{tabular}} & IT-FGSM & 0.06159 & 11.73 & 17 \\ \cline{2-5} 
 & Opt & 0.10455 & 9.63 & 19 \\ \cline{2-5} 
 & Opt\_uni & 0.00005 & 1.73 & 1 \\ \cline{2-5} 
 & AdvGAN & 0.00568 & 3.09 & 1 \\ \cline{2-5} 
 & AdvGAN\_uni & -0.00007 & 2.10 & 1 \\ \hline
\multirow{5}{*}{\begin{tabular}[c]{@{}c@{}}VGG16\end{tabular}} & IT-FGSM & 0.06159 & 9.66 & 35 \\ \cline{2-5} 
 & Opt & 0.24364 & 18.13 & 35 \\ \cline{2-5} 
 & Opt\_uni & 0.00005 & 0.31 & 1 \\ \cline{2-5} 
 & AdvGAN & 0.00568 & 0.75 & 1 \\ \cline{2-5} 
 & AdvGAN\_uni & -0.00007 & 0.44 & 1 \\ \hline

\end{tabular}
}%

\end{table}

\begin{figure}[!t]
\centering
\includegraphics[width = 0.4\textwidth]{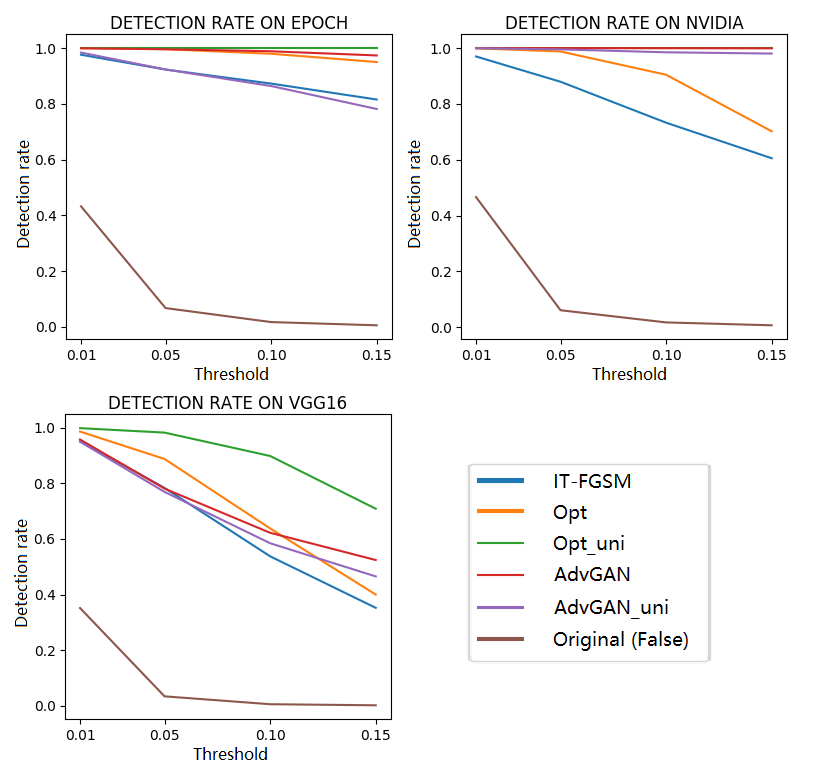}
\caption{Detection rate of Feature Squeezing}
\label{exp3-4}
\vspace{-8pt}
\end{figure}

Table~\ref{exp3-1} shows the attack success rate of each attack method when defended by adversarial training. Figure~\ref{exp3-2} shows the defense result of defensive distillation. For adversarial training, each driving model is re-trained on five new datasets augmented with adversarial examples generated by one of the five attacks. For defensive distillation, We distill seven new models using different temperature values \(\lambda\). The original model is denoted by \(\lambda = 0\). We find that both defense methods are to some extent effective to IT-FGSM. For example, when defending the Epoch model, adversarial training reduces the success rate of IT-FGSM from 59.2\% to 39.3\%, while defense distillation reduces it from 60\% to near 10\% when setting \(\lambda\) to $0.1$. Besides, defensive distillation is also effective in reducing the attack success rate of Opt. However, two defenses are not effective to defend other attacks. 
In particular, defensive distillation essentially smooths the change of gradients in a neural network model to reduce the influence of adversarial perturbations. This explains why it could work on IT-FGSM and Opt but is not effective in other attacks like AdvGAN, which generates adversarial perturbations by learning important features that may affect prediction results.

Table \ref{exp3-3} shows the prediction time overhead (delay), GPU memory overhead, and GPU utilization overhead induced by different adversarial attacks. 
Only IT-FGSM and Opt induces some latency, while other methods induce negligible delays. 
Regarding GPU memory usage, both IT-FGSM and Opt cause over 50\% overhead when attacking Epoch. The memory usage overhead is relatively lower but still significant (around 10\%) when attacking Nvidia and VGG16. Regarding the GPU utilization rate, both IT-FGSM and Opt cause an average of 35\% overhead on Epoch and VGG16 and about 19\% on Nvidia. We only observe 1\% overhead when running the other three attacks. Overall, IT-FGSM and Opt attacks are more likely to be detected because both of them require a few iterations to compute gradients for each input image, which leads to computation overhead. Adv\_GAN only increases GPU utilization by 1\% but increases the memory usage by 4\% for a relatively small model like Epoch. Since Opt\_uni AdvGAN\_uni apply a universal perturbation on each image, they barely cause overhead. Therefore, Opt\_uni and AdvGAN\_uni are less likely to be detected by  runtime monitoring utilities adopted by automotive industry.

Figure \ref{exp3-4} presents the defense effect of feature squeezing. The \textit{Original (False)} line denotes the false detection rate on the original dataset without attack. The other five lines denote the detection rates of adversarial examples for the driving model. When the threshold is set to $0.01$, feature squeezing can almost detect all adversarial attacks on three models. However, it leads to many false positives ($40\%$), which is not realistic to be deployed in practice. When the threshold is set above $0.05$, the effectiveness of feature squeezing decreases significantly. 
Overall, $0.05$ is empirically the best threshold, with 78\% attack detection rate 
and 
only 5\% false positives. 
We also notice when the threshold is larger than $0.01$, the detection rates on VGG16 are lower and decrease faster compared with Epoch and Nvidia.  The result suggests that adversarial examples generated on VGG16 are more resistant to feature squeezing. This is also consistent with our previous finding that adversarial examples generated on more complex regression models are more dangerous. 

\begin{framed}
\noindent Result 3:  
Adversarial training and defensive distillation are effective against IT-FGSM and the optimization based attack to some extent, but no other attacks. Anomaly detection and feature squeezing mechanisms, on the other hand, are able to detect more adversarial attacks but with their own shortcomings.

\end{framed}

\section{Discussion}

\subsection{Implications}


    
\noindent{\bf\em Implication 1. It is important to apply multiple defense methods in combination.}
 We find that no single defense method can effectively protect driving models from all of five attack methods we investigated.
Though feature squeezing has relatively high attack detection rate for all attacks, it has mistakenly detected normal input images as adversarial examples with up to 40\% false positive rate. Thus, these defense methods should be combined to provide a more robust defense against various attacks. 
For some detection methods that require higher computational resources, these techniques might not be fit to be implemented on the vehicle's side. In such a case, we can leverage edge computing for faster response and implement adversarial detection middleware for autonomous driving on edge nodes to enhance autonomous driving vehicle safety.



\noindent{\bf\em Implication 2. 
Further investigation is needed to explore the impact of different DNN structures of regression models on their vulnerability.
} 
We find that VGG16 is less vulnerable than the other two models in both the white-box and black-box settings. 
One possible reason is that VGG16 has more complicated model structure than the other two models, which makes it harder to attack. 
Therefore, it may be more secure to deploy models with complex structures. However, since autonomous vehicles cannot carry super complicated driving models due to limited computation power, we should explore how to design a model with a proper structure complexity while consuming minimal computation power. 
One promising direction is to adopt edging computing architecture~\cite{li2018learning} to deploy such complex driving models. Complex models can be segmented and distributed to end nodes (vehicles), edge nodes (roadside units) and cloud~\cite{han2019convergence} to reduce the resource consumption to vehicles. 

\noindent{\bf\em Implication 3. It is important to protect the detail of driving models and exploring 
the transferability of adversarial examples on driving models is needed.
} The experiment result of the white-box attack (RQ1) shows that attacks 
achieve high attack success rates on driving models, which means once the details of driving models are known by attackers, they could easily implement effective adversarial attacks. Therefore, it is important to hide 
the neural network structure details and hyper-parameters of driving models. On the other hand, recent research shows that deep learning models are susceptible to information extraction~\cite{tramer2016stealing,oh2019towards} and the extracted information can be
used to construct attacks. Therefore, it is also 
important to research 
techniques like~\cite{juuti2019prada} to protect deep learning models against model extraction.
The experiment result of black-box attack (RQ2) shows that adversarial examples generated on a driving model do not have good attack transferability on other driving models. This finding contradicts with existing findings of attack transferability on classification models~\cite{liu2016delving, papernot2016transferability, papernot2017practical}. Since driving models are essentially regression models, this suggests further analysis to investigate the root cause of transferability differences between classification models and regression models. In~\ref{sec:rq2}, we hypothesize that different regression models may learn different hyper-planes to fit data. This hypothesis needs to be evaluated in future work. Future research of transferability of driving models attack will provide more insights about how to construct and defend black-box attacks on driving models.
\subsection{Limitations}

This work assumes attackers could intrude an autonomous driving system to inject malware when an autonomous vehicle is connected to the Internet and upgrades its software and firmware over the air (OTA)~\cite{othmane2015survey}. Then attackers could use the malware to intercept images and construct adversarial examples before the images are fed into the perception layer. Unlike the direct interference to vehicle devices, which could be detected and prohibited by the autonomous driving system and human drivers, adversarial examples are imperceptible to human eyes and thus cannot be easily identified by drivers from the infotainment system in a car. Therefore, when an attacker intrudes a driving system, an simple attack like replacing driving scenes with arbitrary images will be detected easily while adversarial attacks cannot be detected by infotainment system. 

This work only experimented with three CNN based driving models. We did not select the top 3 driving models on the Udacity leaderboard, since all of them take driving videos as input and use sequence-to-sequence structures that existing attack and defense methods are not applicable to.
Furthermore, we have not experimented with driving models with more complicated architectures such as CNN+RNN. 
Similar to prior work~\cite{zhang2018deeproad,tian2018deeptest}, our experiments are only conducted on the Udacity dataset. Validating our findings with more driving models and datasets remains as future work.  

CleverHans \cite{papernot2016cleverhans}, and Foolbox \cite{rauber2017foolbox} are two common open source tools to perform adversarial attacks and defenses. However, these three tools all focus on classification models. Investigating how to adapt those tools to handle regression models that predicate steering angles remains as future work.

In~\cite{he2017adversarial}, an optimization-based approach is proposed to bypass feature squeezing but this approach needs to consume much more time. On the MNIST dataset, this approach spends $20-30$ seconds for generating one valid adversarial example. This time consumption overhead is unrealistic for attacking driving models in real-time so we choose not to implement this approach. 

\section{Conclusion}

This paper presents a comprehensive analysis of adversarial attacks and defenses on autonomous driving models. 
To that end, we implemented five adversarial attacks and four defensive techniques on three CNN based driving models. From experiment results, all of these three driving models are not robust against these adversarial attacks apart from IT-FGSM, while none of four defensive techniques can defend all of five adversarial attacks.
We also raise several insights for future research including 
building middleware that leverages multiple defense methods in tandem, leveraging distributed complex regression driving models, and techniques to defend model information extraction.

\section{Acknowledgments}
We would like to thank anonymous reviewers for their valuable feedback. This work is in part supported by NSF grants CCF1764077, CCF-1527923, CCF-1723773, ONR grant N00014- 18-1-2037, Intel CAPA grant, and Samsung grant.

\clearpage
\bibliographystyle{plain}
\bibliography{reference.bib}

\end{document}